\documentclass[12pt]{article}
\usepackage{graphicx}
\begin{document}
\centerline{\bf \large  On the wave mechanics of a particle in two different} 

\smallskip
\centerline{\bf \large impenetrable spherical cavities}

\bigskip
\centerline{\bf Samrat Dey and Yatendra S. Jain}

\smallskip
\centerline{Department of Physics, North-Eastern Hill University,}

\smallskip
\centerline{Shillong-794022, Meghalaya, India}

\vspace{2.1cm}

\begin{abstract}
Wave mechanics of a particle in an impenetrable spherical cavity  
{\it with} and {\it without} a hard core scattering potential of size 
$\epsilon$ ($\epsilon \to 0$) at its 
center is critically analyzed.  It makes marginal but important 
modifications in our present understanding of the energy eigen values of 
such particles.   
\end{abstract}


\vspace{1.1cm}
\noindent
\copyright with authors.
\vspace{0.9cm}

\bigskip
\centerline{\bf 1. Introduction}

\bigskip
Wave mechanics of a particle, placed in widely different potentials and 
confinements ({\it e.g.}, 1-D, \, 2-D \, and \, 3-D boxes, spherical 
cavity, {\it etc.}), has been a subject of great significance$^{1-6}$ 
and persisting investigation$^{7-11}$ for various reasons.  For example, 
the study of a particle in a spherically symmetric potential $V(r)$ helps 
in understanding several physical systems such as hydrogen atom$^{1}$, 
atomic nucleus$^{10, 12}$, electron bubble$^{13, 14}$, quantum scattering 
of particles$^{1, 2, 15}$, {\it etc}.  In this paper, we make a critical 
analysis of two separate spherical impenetrable cavities [say, Cavity-(i) 
and Cavity-(ii) of radii ${\rm R}$ with {\it centers identified by 
point} ``C''] with a point particle, trapped inside. Each of these cavities 
within its bounds has zero potential at all points except at C of 
Cavity-(i) which is assumed to be occupied by a hard core sphere 
(of radius $\epsilon \sim 0^+$, -a little more than $0$) representing a 
 $V(r)$, defined by 
$V(r \leq \epsilon) = \infty$ and $V(r > \epsilon) = 0$.  
We note that the eigen energy solutions for a particle constrained to 
move under the influence of a $V(r)$ are obtained by solving the 
radial part of its 3-D Schr\"{o}dinger equation  
$$\frac{1}{r^2}\frac{d}{dr}\left(r^2\frac{dR_{l}(r)}{dr}\right) + 
\left[\frac{2M}{\hbar^2}(E-V(r)) - \frac{l(l+1)}{r^2}\right]R_{l}(r)
 = 0. \eqno(1)$$

\noindent 
Here, $M=$ mass of the 
particle, $\hbar = h/2\pi$ (with $h$ being the 
Planck's constant) and $R_l(r)$ (with $l$ representing orbital angular 
momentum of the state of the particle) is an eigen function of eigen 
energy $E$.  Although, as defined above, Cavity-(i) and Cavity-(ii) 
are physically different systems for the potential they offer to 
the motion of the trapped particle, the known energy eigen 
values, $E_{n,l}(i)$ ($n$ being the principal quantum number) of 
the particle in Cavity-(i), obtainable from$^{16}$ ({\it cf.}, Eqn.5 
and its generalization for all $l$, as concluded in 
Section 2.6)
and $E_{n,l}(ii)$ 
of the particle in Cavity-(ii), 
available from several graduate texts$^{3-5}$, satisfy 
$E_{n,l}(i) = E_{n,l}(ii)$ ({\it cf.}, Eqn.8 and Section 2.6).  
This {\it unexpected equality} motivated us to investigate its origin 
and to suggest correct values of $E_{n,l}(ii)$.  

\bigskip
\centerline{\bf 2. Analysis and Discussion}

\bigskip
\noindent
{\it 2.1 Unexpected Equality of $E_{n,0}(i)$ and $E_{n,0}(ii)$} 

\bigskip
Eqn.1 has been solved by Huang and Yang$^{16}$ for a particle trapped 
in the space between two concentric impenetrable spherical surfaces 
of radii ${\rm R}$ and $a$ (with ${\rm R} > a$).  Using the boundary 
condition, $R_{l=0}(r = {\rm R}) = R_{l=0}(r = a) = 0$, they obtained  
$$R_{0}(r) = B\frac{\sin{k(r-a)}}{r} \quad {\rm for} \quad a<r<{\rm R} 
\quad {\rm and} \quad \quad R_o(r\le a) = 0 \eqno(2)$$

\noindent
where $B$ stands for normalization constant and 
$$k_{n,l=0} = \frac {\pi n} {({\rm R}-a)}, \quad \quad n = 1, 2, 3, 
...\eqno(3)$$

\noindent
which represent the allowed values of momentum wave vector $k$.  
Using $a = \epsilon$, we have 
$$k_{n,0} = \frac{\pi n}{{\rm R - \epsilon}} =  \frac{\pi n}{{\rm R}},
\quad {\rm as} \quad {\rm R} >> \epsilon
 \quad \quad n = 1, 2, 3, ..., \eqno(4)$$
$$E_{n,0}(i) = \frac{n^2h^2}{8M{\rm R}^2} \eqno(5)$$

\noindent
and 
$$R_{0}(r) = B\frac{\sin{[k(r-\epsilon)]}}{r} \approx B\frac{\sin{kr}}{r} 
\quad {\rm at} \quad r > \epsilon \quad {\rm and} \quad \quad R_o(r\le
\epsilon) = 0   \eqno(6)$$

\noindent
for the particle in Cavity-(i).  However, what follows from$^{3, 4}$, we 
also have  
$$E_{n,0}(ii) = \frac{n^2h^2}{8M\rm R^2} \eqno(7) $$

\noindent  
which can be compared with Eqn.5 to discover an equality,       
$$E_{n,0}(i) = E_{n,0}(ii),  \eqno(8)$$

\noindent
not expected for a particle placed in two physically different 
cavities.  This underlines the problem that we intend to resolve 
in this paper.  
\clearpage
\bigskip
\noindent
{\it 2.2 Equivalence of Plane Wave and Spherical Waves} 

\bigskip
When the Schr\"{o}dinger equation for a free particle, expressed in 
spherical polar coordinates, is solved one finds that different 
states of the particle are described by spherical waves$^{3, 4}$ 
({\it centered around} $r=0$ of the coordinate system) whose 
radial part is expressed by 
$$R_l(r) = A_lj_l(kr) \eqno(9)$$

\noindent
which represents the solution of Eqn.1 with $V(r) = 0$; 
here $A_l$ are normalization constants and $j_l(kr)$ are spherical 
Bessel functions.  However, when we solve the Schr\"{o}dinger 
equation expressed in Cartesian coordinates, we find that the same 
particle is described$^{1, 2}$ by a plane wave, 
$$u_{\bf k}({\bf r}) = e^{i{\bf k}.{\bf r}} \eqno(10)$$

\noindent
which is presumed to have unit normalization; here {\bf p} = {$\hbar$\bf k} 
and {\bf r}, respectively, represent momentum and position vector 
of the particle.  This indicates an obvious {\it inter-relationship} 
between the two descriptions of a free particle and in what follows 
from$^{1-6}$, we have  
$$e^{i{\bf k}.{\bf r}} = \sum_{l=0}^{\infty}\sum_{m=-l}^{l}
R_l(r)Y_{lm}(\theta, \phi), \eqno(11)$$

\noindent
for a particle moving in any general direction (here 
$Y_{lm}(\theta, \phi)$ are spherical harmonics) and 
$$e^{i{k}{z}} = \sum_{l=0}^{\infty} R_l(r)P_{l}(\theta) \eqno(12)$$

\noindent
for a particle moving in $z-$direction (here $P_{l}(\theta)$ are 
associated Legendre polynomials).  One uses$^{3,4}$ certain boundary 
conditions on $R_l(r)$ to determine $E_{n.l}(ii)$ and, in doing so, 
it is implicitly assumed that spherical waves, $R_l(r)$, originate at the 
center, C, of Cavity-(ii) which, however, violates cause and 
effect relationship ({\it cf.} Section 2.3).     

\bigskip
\noindent
{\it 2.3 Physical relevence of Eqn.11}

\bigskip
We have no doubt that Eqn.11 (or Eqn.12) is a mathematically sound relation, 
but it is important to emphasize that a plane wave [$u_{\bf k}({\bf r})$, 
Fig.1(a)] would transform into spherical waves only when it meets a 
scatterer [{\it a source of scattering potential}, $V(r)$] and 
this inference is corroborated by the experimental fact that 
neither a ray of light, nor a beam of particles, is observed 
to have its scattering around an arbitrary point on its path;  
{\it the point should be occupied by a scatterer} as a source 
of $V(r)$.  To understand this, it may be noted that 
the transformation of a plane wave (Fig. 1(a)) into spherical 
waves (Fig.1(b) physically means that the incoming particles
represented by the plane wave  
deviate from their paths and turn to move along any of the infinite 
many arrows having their single cross point ($r=0$) 
[{\it cf.}, Figs.1(c)] which also represents the center of 
spherical waves [{\it cf.}, Fig.1(b)].  
However, as a {\it law of nature}, such a deviation is possible 
only if the cross point of the arrows [Fig.1(c)] is occupied by 
a source of potential $V(r)$ which can serve as the origin of 
a deviating force.  In fact, it is for this reason that 
the partial wave theory of 
scattering of a beam of particles uses Eqn.12 to convert an 
incoming plane wave into spherical waves only around the location 
of a scatterer ({\it and not around any arbitrary point}) and succeeds 
reasonably well in accounting for the experimentally observed scattering 
cross sections$^{1-6}$ and other aspects.  

\bigskip
In this context, it may be noted that if a 
single particle  happens to repeatedly meet a $V(r)$ (for its physical 
situations such as confinement in a
spherical cavity) and get scattered in 
different directions [say along infinitely many possible lines passing 
through the center $r=0$ of $V(r)$], it would also render an overall 
result which could be expressed in terms of spherical waves. 
This explains how a state of a single particle ({\it trapped in a 
spherical cavity }), which at any instant has a linear motion before 
and after an event of collision with a scatterer [be it on the surface 
or on at C, in case of Cavity-(i)], can be identified with 
spherical waves, originating from the scatterer, 
when one sees collective result of infinite many 
collisions with the scatterer. 
This account is consistent with our understanding of the interference pattern 
obtained from a double slit interferometer in which only one electron is 
allowed to pass through its slits at a given time but the experiment is
performed by using very large number of monoenergetic electrons$^{17}$.  
While, each electron
hitting the viewer's screen leaves its record only at one point, the 
interference pattern emerges as a culmination of a very large number of 
points where different electrons hit the screen one by one$^{17}$. 

\bigskip
A hard sphere of radius $\epsilon$ [{\it i.e.} a $V(r)$], presumed to occupy C of Cavity-(i), renders a 
clear cause for the scattering of the trapped particle and hence, 
for the transformation of a plane wave into spherical waves 
at this point.  Evidently, the spherical waves ({\it representing 
the scattering of particle}) so produced, have to have their 
center at the said C point; in other words $r=0$ of $R_l(r)$ of 
these waves is located at the C of Cavity-(i).  However, the 
particle in Cavity-(ii) has no reason to get scattered around 
its C, because it finds no source of scattering potential at this 
point.  Naturally, the implicit assumption (behind $E_{n,l}(ii)$, 
-reported in$^{3,4}$) that a plane wave representing the trapped 
particle can burst into spherical waves at the C of Cavity-(ii), is incorrect.  In what follows from these observations, the center $r=0$ 
of spherical waves ($R_l(r)$), representing the scattering of 
particles, falls {\it only at} the $r=0$ of scattering potential 
$V(r)$ and this represents an important aspect for the 
transformation of a plane wave into spherical waves in the 
domain of physics.

\bigskip
\noindent
{\it 2.4 Meaning of $R_l(r)$ and $\chi_l(r)$}

\bigskip
We note that Eqn.1 can be transformed into   
$$\frac{d^2\chi_l(r)}{dr^2} + \left[\frac{2M}{\hbar^2}
(E-V(r)) - \frac{l(l+1)}{r^2}\right]\chi_l(r) = 0, \eqno(13)$$

\noindent
by using
$$R_l(r) = \frac{\chi_l(r)}{r}. \eqno(14)$$

\noindent
While, this means that solutions of Eqn.1 can be obtained by solving 
Eqn.13 for its eigen energy $E$ and corresponding $\chi_l(r)$, it 
also helps in finding the meaning of $R_l(r)$ in relation to that 
of $\chi_l(r)$.  $\chi_l(r)$ represents radial probability of the particle along a line passing 
through the scatterer, where as, $R_l(r)$ (radial part of the spherical wave) 
represents the distribution of the said linear probability ({\it with equal probability }) over other 
identical lines ({\it infinitely large in number}).  This can 
be clarified by using $|R_l(r)| = |\chi_l(r)|/\sqrt{4\pi}r$, 
a slightly modified form of Eqn.14 which renders   
$$|R_l(r)|^2  = \frac{|\chi_l(r)|^2}{4\pi r^2}. \eqno(15)$$

\noindent
Since, the said modification involves only a constant factor, $1/\sqrt{4\pi}$, 
Eqn.15 is consistent with Eqn.13. It may, therefore, 
be concluded that while, $\chi_l(r)$ represents the 
amplitude of the radial probability of finding the particle at a point on a 
radial line at a distance $r$ from the scatterer, $R_l(r)$ represents the 
amplitude of the uniform distribution of the probability $|\chi_l(r)|^2$ 
over all points on $4\pi r^2$ surface area of the shell of radius $r$ 
around the scatterer.  In other words $r$ appearing in the denominator 
of Eqn.14 signifies only this distribution and $\chi_l(r)$ carries 
full information about the physical state the particle and allowed 
energy values. Evidently, it is sufficient to put the one dimensional boundary condition on $\chi_l(r)$, the representation of the radial distibution along a line.  

\bigskip
\noindent
{\it 2.5  Nature of Waves in Cavity-(i) and Cavity-(ii)}

\bigskip
The hard core potential in Cavity-(i) scatters the particle around 
its C and produces spherical waves centered at this point.  These 
waves can be seen to have two parts of their journey: (Part-1) 
originating from C, they move away to strick on the cavity 
surface, and (Part-2) reflected back from the said surface, they 
converge at C from where they turn back to repeat the cycle of 
Part-1 and Part-2.
However, as cavity-(ii) offers no scattering potential to the trapped 
particle at its center C, spherical waves are not expected to 
originate from this point.  To believe that the particle in 
Cavity-(ii) assumes a state represented by spherical waves, we use 
the fact that the surfaces of Cavity-(ii) and Cavity-(i) 
offer a potential of {\it identical form} to scatter/reflect the 
particle.  Naturally, spherical waves for the particle in 
Cavity-(ii) can originate from its surface, in a manner such 
waves (Part-2) are assumed to originate from the surface of 
Cavity-(i); to understand the said manner, it may be noted 
that a particle colliding at any point on the surface is likely 
to get scattered along any of the infinite many arrows passing 
through this point.  Since all points on the inner surface 
are identical, the particle gets identically scattered from 
them and the net effect is a spherical wave front [Fig. 1(d) 
shows how it could be constructed from the spherical 
wavelets emerging from all such points] which moves towards 
the C of the cavity.  While in Cavity-(i) this wave front 
is scattered back to move toward the cavity surface, it is natural 
this wave front in Cavity-(ii) crosses the center C without 
getting scattered from it [{\it in contrast to their obvious scattering} 
from the C of Cavity-(i)] and it is diverted back only 
after its touch with the cavity surface.  It is clear 
that the wave function, representing any state of the particle 
in Cavity-(ii), is not forced to vanish at its center C as it 
happens for the state of the particle in Cavity-(i).        
This, naturally, means that $r=0$ of $R_l(r)$  of these waves 
in Cavity-(ii) lies at its surface  
(where there are, indeed, scattering potentials) not at its C.  

\bigskip
\noindent
{\it 2.6 Eigen Values of the Particle in Cavity-(i) and Cavity-(ii)}

\bigskip
Now, having a clear qualitative picture of the nature of the waves 
in these two cavities, let us calculate the energy eigen values.  
Considering that each point on the spherical wall of these cavities 
is a center ($r=0$) of strongly repulsive potential $V(r)$, we find 
that on {\it any radial line} in Cavity-(i), scattering 
potentials are present with their centers at 
{\it three} points, (see three dark spherical dots marked as 1, 2, and 3 
in Fig. 1(e)), while, in Cavity-(ii) they are present only at {\it two} 
points (see two dark spherical dots marked as 1 and 2 in  Fig. 1(f)). 
It is common to observe a single system having several 
centers of scattering potential(s), for example, numerous constituents 
(atoms/molecules) 
of a solid serve as centers the scattering of x-ray photons, electrons, 
neutrons, {\it etc.}.  

\bigskip
In what follows from Section 2.3, the origin of spherical waves is 
synonym with the center $r=0$ of $V(r)$, and Section 2.4, $\chi_l(r)$, subjected 
to appropriate boundary
conditions, is sufficient to determine the energy eigen values of the particle. 
Guided by these observations, we choose the origin 
$r=0$ of $\chi_l(r)$ {\it only} at any of the three points for 
Cavity-(i) [{\it viz.} points 1, 2 and 3 in Fig.1(e)] and any of 
the two points for Cavity-(ii) [{\it viz.} points 1 and 2 in Fig. 1(f)] 
from where the scattering of the particle can take place. In this 
context, as concluded above, $R_l(r)$ has zero value at the location 
$r=0$ (even for $l = 0$ ({\it cf.} Eqn.6) which implies that $\chi_l(r)$ is also zero at 
this location.  For an identical choice for both cavities, we 
prefer to choose point 1 (or 2) [Figs.1(e) and 1(f)] as the 
center of $V(r)$ and the origin of $\chi_l(r)$ and use the location 
of nearest scattering center on the radial line as a point where 
$\chi_l(r)$ has to have its node (zero value); however, for Cavity-(i), 
one may also choose point 3 as the origin $r=0$ of $\chi_l(r)$ and 
point 1 (or 2) as a location where it has to have a node. Evidently, 
the nodes of $\chi_l(r)$ are separated at the maximum by ${\rm R}$ in 
case of Cavity-(i) and by $D (= 2 \rm R)$ in case of Cavity-(ii).

\bigskip
Since, $\chi_l(r)= rR_l(r) = rA_lj_l(kr)$ and $j_l(kr)$ has many 
zeroes at $kr = \beta_{n,l}\pi$ (where $\beta_{n,l}$
is a number as defined in$^3$), the energy eigen values for the 
particle in Cavity-(i) are given by
$$E_{n,l}(i) = \frac{\beta^2_{n,l}h^2}{8M{\rm R}^2} \eqno(16)$$

\noindent
by setting 
$$k_{n,l}{\rm R} = \beta_{n,l}\pi, \eqno(17)$$

\noindent
and those in Cavity-(ii) are given by 
$$E_{n,l}(ii) = \frac{\beta^2_{n,l}h^2}{8MD^2} \eqno(18)$$

\noindent
by setting
$$k_{n,l}D  = \beta_{n,l}\pi. \eqno(19)$$

\bigskip
We now note that our $E_{n,l}(i)$ (Eqn.16) not only encompasses Eqn.5, 
but also generalizes Eqn.8 for all values of $l$ as it is equal to  
$E_{n,l}(ii)={\beta^2_{n,l}h^2}/{8M{\rm R}^2}$ concluded in$^{3, 4}$.  However,
$E_{n,l}(ii)$ (Eqn.18), concluded from this study, differs from 
$E_{n,l}(ii)$ of$^{3, 4}$ by a factor of 1/4.   It is because, the 
steps of finding 
$E_{n,l}(ii)$ in$^{3, 4}$ somehow presume that $u_{\bf k}({\bf r})$ 
of the particle bursts into spherical waves at C of Cavity-(ii), 
although necessary $V(r)$ does not exist there (as discussed in section 2.3); this explains why 
$E_{n,l}(ii)$, reported in$^{3, 4}$, exactly matches with our $E_{n,l}(i)$ 
(Eqn.16) for Cavity-(i) which, indeed, has a infinitely small hard core 
scattering potential at its center. Since, the present 
analysis does not make the said presumption, it renders $E_{n,l}(ii)$ (Eqn.18) 
which differs from $E_{n,l}(i)$ (Eqn.16), as expected.

\bigskip
\centerline{\bf 3. Conclusion}

\bigskip
As discussed in section 2.3, we find that the use of Eqn.11 
(or, Eqn.12) in obtaining $E_{n,l}(ii) = {\beta^2_{n,l}h^2}/{8M{\rm R}^2}$ 
(as concluded in$^{3,4}$) is not proper.  Further, it is for this reason, 
that one finds an unrealistic equality between $E(i)$ and $E(ii)$ 
({\it cf.}, Eqn.8).  Guided by this fact, we analysed the problem more 
critically and discovered that $E_{n,l}(ii)$ (Eqn.18) is clearly different 
from  $E_{n,l}(i)$ (Eqn.16) by a factor of 1/4.  We hope that this would 
greatly help in having a better understanding of the quantum states of 
a particle trapped in Cavity-(i) and Cavity-(ii).

\bigskip
\noindent
{\it Acknowledgment}: Authors are thankful to Dr. Simanta Chutia and Jeeban
P.  Gewali for useful interaction. The work is supported partially by 
DST project.

\clearpage
\bigskip
\noindent
\centerline{\bf  References}
 
\bigskip
\parindent=-0.15cm
{$^1$}L. I. Schiff, {\it Quantum Mechanics}, (McGraw Hill, Singapore, 1968).

\bigskip
\parindent=-0.15cm
{$^2$}E. Merbezbacher, {\it Quantum Mechanics}, (John Wiley \& Sons, Singapore, 2004).

\bigskip
\parindent=-0.15cm
{$^3$}S. Flugge, {\it Practical Quantum Mechanics}, (Springer-Verlag, New York, 1974).

\bigskip
\parindent=-0.15cm
{$^4$}G. Arfken, {\it Mathematical Methods for Physicists}, (Academic Press, Inc., London, 1985).

\bigskip
\parindent=-0.15cm
{$^5$}W. A. Harrison, {\it Applied Quantum Mechanics}, (World Scientific, Singapore, 2000).

\bigskip
\parindent=-0.15cm
{$^6$}G. Aruldas, {\it quantum Mechanics}, (Prentice-Hall of India, Delhi, 2002).

\bigskip
\parindent=-0.15cm
{$^7$}R. H. Lambert, Am. J. Phys. {\bf 36}, 417 (1968).

\bigskip
\parindent=-0.15cm
{$^8$}J. Gea-banacloche, Am. J. Phys. {\bf 70}, 307 (2002). 
 
\bigskip
\parindent=-0.15cm
{$^9$}Y. S. Jain, "Physical behavior of a system representing a particle
trapped in a box having flexing size" arXiv: 0807.0732 and several 
references cited therein. 

\bigskip
\parindent=-0.25cm
{$^{10}$}V. G. Gueorguiev and J. P. Draayer, Mixed-Symmetry Shell Model
Calculations, arXiv: nucl-th/0210034v1 (2002); V. G. Gueorguiev Mixed-Symmetry 
Shell Model Calculations in Nuclear Physics, 
Ph.D. Dissertation, Department of physics and Astronomy, M. S. Sofia 
University, December 1992.

\bigskip
\parindent=-0.25cm
{$^{11}$}M. Carreau, E. Farhi and S. Gutmann, Phys. Rev. D. {\bf 42}, 
1194 (1990).

\bigskip
\parindent=-0.25cm
{$^{12}$}M. S. Rogalski and S. B. Palmer, {\it Quantum Physics}, (Gordon and 
Breach Science Publishers, Singapore, 1999), p 397.

\bigskip
\parindent=-0.25cm
{$^{13}$}J. Classen, C.-K. Su, M. Mohazzab and H. J. Maus, 
J. Low Temp. Phys. {\bf 110}, 431 (1998).

\bigskip
\parindent=-0.25cm
{$^{14}$}H. Maris and S. Balibar, Physics Today, (Feb 2000), pp 29-34. 

\bigskip
\parindent=-0.25cm
{$^{15}$}B. H. Bransden and C. J. Joachain, {\it Physics Of Atoms And
  Molecules}, (Pearson Education, Singapore, 2003).

\bigskip
\parindent=-0.25cm
{$^{16}$}K. Huang and C. N. Yang, Phys. Rev. {\bf 105}, 767 (1957).

\bigskip
\parindent=-0.25cm
{$^{17}$}A. Tonomura, J. Endo, T. Matsuda and T. Kawasaki, 
Am. J. Phys. {\bf 57}, 117 (1957).

\clearpage 
\begin{figure}[H]
\begin{center}
\includegraphics[angle = 0, width=.8\textwidth]{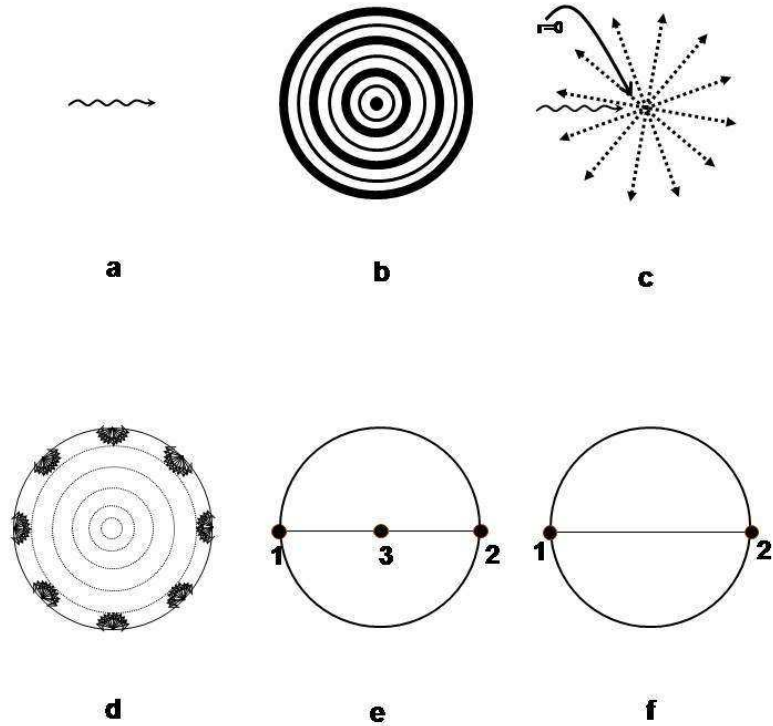}
\caption{ Depiction of: (a) $u_{\bf k}({\bf r})$, (b) 
spherical waves representing 
scattered waves from $V(r)$ centered around $r=0$, (c) some of the 
possible directions of the motion of a particle after its scattering form 
$V(r)$ centered around $r=0$, (d) spherical wave front (dotted circules) 
constructed from scattered wavelets (represented by bunches of arrows 
emerging from different points on the surface) 
from different points on the surface, (e) one of the 
infinite many radial lines of Cavity-(i) and three centers 
(dark spherical dots) of the 
scattering potentials  present on it and (f) one of the infinite many 
radial lines of Cavity-(ii) and two centers (dark spherical dots) 
of the scattering potentials present on it.}
\end{center}
\end{figure}

\end{document}